\documentclass[runningheads]{llncs}

\pdfoutput=1
\usepackage[T1]{fontenc}
\usepackage{graphicx}
\newcommand{\hide}[1]{}
\newcommand{\remove}[1]{}

\begin{document}

\title{Parallel and Distributed Data Series Processing on Modern and Emerging Hardware\thanks{Supported by 
the Hellenic Foundation for Research and Innovation (HFRI) under the ``Second Call for HFRI Research 
Projects to support Faculty Members and Researchers'' (project number: 3684).}}
\author{Panagiota Fatourou\inst{1,2}}
\authorrunning{P. Fatourou}
\institute{Foundation for Research and Technology, Institute of Computer Science, Greece \and
University of Crete, Department of Computer Science, Greece \\
\email{faturu@csd.uoc.gr}}
%\url{http://www.csd.uoc.gr/~faturu}}
%
\maketitle              % typeset the header of the contribution
\begin{abstract}
This paper summarizes state-of-the-art results on data series processing with the
empahsis on parallel and distributed data series indexes that
exploit the computational power of modern computing platforms. 
The paper comprises a summary of the tutorial the author delivered at 
the 15th International Conference on Management of Digital EcoSystems (MEDES'23).
%\keywords{Data series \and Time series \and Indexing \and Similarity search \and Query answering \and Multi-core architectures \and Parallelization \and GPU processing \and Disk-based index \and In-memory index \and Distributed processing.}
\end{abstract}
\section{Introduction}
Ordered sequences of data points, known as {\em data series}, are one of the most common types of data. 
They are heavily produced by several applications in many different domains, including finance, astrophysics, telecommunications,
environmental sciences, engineering, multimedia, neuroscience, 
and many others~\cite{DBLP:journals/sigmod/Palpanas15,fulfillingtheneed}).
%~\cite{kashino1999time,ye2009time,shasha1999tuning,huijse2014computational,raza2015practical}), 
Similarity search is a fundamental building block in data series processing,
as many types of complex analytics, such as, clustering, classification, 
motif and outlier detection, etc., are heavily dependent on it.
A similarity search query is also useful in itself.
It searches for the data seires in the collection 
that has the smallest distance to the query series, given some distance metric
(which can be e.g., Euclidean distance~\cite{Agrawal1993} or Dynamic-Time Warping~\cite{rakthanmanon2012searching}).

As data series collections grow larger, 
sequential data series indexing technologies turn out to be inadequate. 
For example, the state-of-the-art such index, ADS$+$~\cite{zoumpatianos2016ads}, 
requires more than 4min to answer a single query on a moderately sized 250GB sequence collection. 
Thus, ADS is inefficient in processing the enormous sequence collections 
that are produced in many domains.

We briefly present a sequence of state-of-the-art indexes 
that take advantage of multiple nodes and modern hardware parallelization 
in each node,
%and incorporate the state-of-the-art techniques 
%in sequence indexing, 
in order to accelerate processing times and achieve scalability.

We focus on a specific family of indexes, namely the {\em iSAX indexes}~\cite{ZoumpatianosIP16,peng2018paris,parisplus,peng2020messi,messijournal,peng2021sing}. 
An iSAX index first computes the {\em iSAX summary}~\cite{iSAX} of each data series in the collection ({\em summarization phase}). 
Then, it builds a tree storing all summaries ({\em tree construction phase}), 
and uses it to answer similarity search queries. 
Given a query $q$, the similarity search algorithm 
first traverses a path of the tree to find a first approximate answer  
to $q$. This approximate answer is called {\em Best-So-Far} ({\em BSF}). 
Then, it traverses the collection of data series to prune those whose summaries have higher distance to 
$q$'s summary than the value stored in BSF. The distance between the summaries of two data series is called {\em lower bound distance}. 
It has been proved that if the lower bound distance of a series DS from q is higher than BSF, 
the real distance of DS from q is also higher than BSF.
Then, DS can be pruned. Finally, the actual (real) distances between each data series that cannot be pruned and $q$
is calculated. 
Whenever a real distance computation results in a lower value than that of BSF, then BSF is updated. Eventually,
the answer to $q$ is the value stored in BSF. 
If the pruning degree is high, a vast amount of expensive real distance computations are avoided.
This results in good performance.

We first discuss ParIS+~\cite{peng2018paris,parisplus}, a {\em disk-based} concurrent index, capable to manipulate big data series collections stored in secondary storage. 
We then continue to present MESSI~\cite{messi,messijournal}, the state-of-the-art {\em in-memory} concurrent index. Next, we discuss the 
challenges originating from the utilization of Graphics Processing Units (GPUs)~\cite{peng2021sing} 
for further improving performance during query answering. We also discuss SING~\cite{peng2021sing},
an index that utilizes GPUs for query answering to perform better than MESSI. 
These indexes improve performance drastically in comparison 
to ADS$+$, the state-of-the-art sequential index~\cite{zoumpatianos2016ads}: to answer 
a similarity search query on a 100GB (in memory) data series collection,
SING requires 35 msec whereas ADS$+$ requires several tens of seconds.  

Finally, we discuss Odyssey~\cite{odyssey}, a {\it distributed} 
data-series (DS) indexing and processing framework,
that efficiently addresses %the enhanced performance 
the challenges for efficient and 
highly-scalable {\it multi-node} data series processing. 

All the indexes that we discuss in the next sections, exploit the Single Instruction Multiple Data (SIMD) capabilities of modern CPUs 
to further parallelize the execution of individual instructions inside each core. 

%The state-of-the-art sequential data series index~\cite{}, known as ADS,
%required +++++ to answer a query series on a collection of 100GB data series. 
%As the collections of data are growing in enormous speeds, much more efficient and scalable solutions are required. 

\section{ParIS+}
\label{sec:paris}
ParIS+~\cite{peng2018paris,parisplus} is a disk-based index that takes advantage of 
the computational power of multi-core 
architectures in order to execute in parallel the computations needed for both index creation and query answering. 
ParIS+ makes careful design choices in the coordination of the computational and I/O tasks,
thus managing to completely hide the CPU cost during index creation under I/O.

ParIS+ is 2.6x faster than ADS$+$~\cite{zoumpatianos2016ads} in index creation,
and up to 1 order of magnitude faster in query answering.
%We also note that ParIS and ParIS+ have the potential to deliver more benefit as we move to faster storage media.
%and that it has the potential to deliver even more benefit when we move 
%to faster storage mediums for dataset storage.
%in order to achieve these goals, we had to make careful design choices in the coordination 
%
This makes ParIS+ a very efficient solution for disk-resident data. %Its performance is dominated by the I/O cost. %s it encounters. 
However, its answering time to a search query on a 100GB dataset is 15sec. 
This performance not only does not support interactive analysis (i.e., 100msec)~\cite{Fekete:2016},
but also it is above the limit for keeping the user's attention.

To move data from disk to main memory, ParIS+ uses a double buffering scheme.
While a coordinator thread moves data to one part of the buffer,
a number of worker threads compute the iSAX summaries for the data series stored into the other part,
and store them in a set of {\em iSAX buffers}. The use of the double buffering scheme enables ParIS+ to hide CPU computation under I/O. 

The worker threads traverse the iSAX buffers
storing pairs of iSAX summaries and the pointers to the corresponding data series, into the leaves 
of the index tree. 
The iSAX buffers are needed to ensure some form of data locality. All data series found in an iSAX buffer  
are stored in the same subtree of the index tree. ParIS+ uses a dedicated thread to build each
subtree. This eliminates the need for costly synchronization and communication between the threads. 

During query answering, ParIS+ first calculates BSF using the index tree (and never uses the tree again). 
Then, a number of threads concurrently traverse different parts of an array, called SAX, 
which contains the iSAX summaries of all data series of the collection, in order to create a list 
({\em candidate list}) of those series that are good candidates to be the query's answer
(i.e., those whose lower bound distance is smaller than BSF and cannot be pruned). 
Finally, a number of threads perform in parallel,
using SIMD, the real-distance computations needed to find the closest data series to the query, among the candidate series. 

\section{MESSI}
\label{sec:messi}
The second index, called MESSI~\cite{messi,messijournal}, provides an efficient indexing and query answering scheme 
for \emph{in-memory} data series processing. 
Fast in-memory data series computations often appear in real scenaria~\cite{Palpanas2019,bigdata20tutorial}.
For instance, Airbus stores petabytes of data series, reasoning about the behavior of aircraft components or pilots~\cite{Airbus},
but requires experts to run analytics only on subsets of the data (e.g., on those relevant to landings 
from Air France pilots) that fit in memory.  
MESSI features a novel solution for answering similarity search queries
which is 6-11x faster than an in-memory version of ParIS+, 
achieving for the first time interactive exact query answering times, at $\sim$60msec.
It also provides redesigned algorithms that lead to a further $\sim$4x speedup in index construction time, in comparison to (in-memory) ParIS+.

The design decisions in ParIS+ were heavily influenced by the fact that 
its performance cost was mostly I/O bounded. %Of course, this cost goes away in its in-memory version. 
Since MESSI copes with in-memory data series, no CPU cost can be hidden under I/O.
Therefore, MESSI required more careful design choices %in the setup 
and coordination of the parallel workers. % when accessing the required data structures, in order to improve its performance. 
This led to the development of a more subtle design for the index construction 
and new algorithms for answering similarity search queries on this index. 

For query answering, in particular, the MESSI paper shows that an in memory version of ParIS+ 
is far from being optimal. Thus, MESSI proposes new solutions to achieve a good balance 
between the amount of communication among the parallel worker threads, and the effectiveness of each individual worker.

MESSI uses concurrent priority queues for storing the data series 
that cannot be pruned, and for processing them in order, starting from those that have the smallest 
lower bound distance to the query series. 
In this way, the worker threads achieve a better degree of pruning.
Moreover, MESSI assigns an iSAX summary to each node of the index tree.
Then, it traverses the tree to decide which data series cannot be pruned based on these iSAX summaries. 
If the lower bound distance of a node from the query series is lower than BSF, the entire subtree
rooted at the node can be pruned.
In this way, the number of lower bound distance calculations performed
is significantly reduced.
To achieve load balancing, MESSI ensures that 
all priority queues have about the same number of elements,
%of data series for which MESSI has to calculate the distance
%between their iSAX summaries and the query summary. 
and workers use randomization to choose the priority queues they will work on.

\section{SING}
\label{sec:sing}
Although ParIS+~\cite{peng2018paris,parisplus} and MESSI~\cite{peng2020messi,messijournal} exhibit
advanced performance by exploiting the parallelism opportunities offered by the multi-core and SIMD architectures,
they ignore the computational power of Graphics Processing Units (GPUs).
SING~\cite{sing} is the first (in-memory) 
data series indexing scheme that combines the GPU's parallelization opportunities with those of multi-core architectures (and SIMD), 
to further accelerate exact similarity search.
SING outperforms MESSI in a variety of settings, reducing the MESSI cost for query answering to (almost) half.
%, 
%even when we use all 16 cores %(32 hyperthreads) 
%of our system. %The results show that SING delivers significant performance benefits in terms of query processing time across the range of datasets and query workloads we tried. 

Data series processing with GPUs is challenging for several reasons. 
First, the GPU memory is rather limited, %and very small when compared to the dataset size.
%Therefore, the GPU 
so storing the entire raw data set in it is impossible.
Moreover,  the %(relatively) 
slow interconnect speeds disallows processing raw data in the GPU memory,
as moving even small subsets of such data (i.e., those data series that are not pruned) %in the summarized space) 
in the GPU memory incurs a prohibitively high cost. 
Last, GPUs employ non-sophisticated cores,
which are not %the ideal computing environment for 
readily suited to 
processing tree indexes and algorithms on top of them that frequently involve branching.
These considerations imply that it is just (parts of) the query answering computation
%existing tree-based indices
that can be performed efficiently in the GPU. Furthermore,
the SING paper provides experimental evidence that simple %extensions 
GPU adaptations of the techniques employed by previously-presented tree-based indices 
for implementing those parts cannot %lead to efficient algorithms, able to 
outperform state-of-the-art solutions, such as MESSI.

To address these challenges, SING provides a new similarity search algorithm that
runs on top of the tree structure created by MESSI.
The algorithm ensures the efficient collaboration of both the CPU and GPU cores
to answer queries. The main ideas on which it is based are the following.
SING stores in the GPU memory, an array of (just) the iSAX summaries
of the data series in the collection. This array is sorted in the order the data series 
appear in the leaves of the index tree. 
SING performs an inital pruning phase to prune entire root subtrees, thus
reducing the number of lower bound distance computations that the GPU executes.
Moreover, it employs a simple polynomial function to compute lower bound distances.
This enables the computation of lower bound distances in their entirety within the GPU.
%and avoids expensive accesses to memory outside the Streaming Processors.
SING employs streaming to effectively synchronize the parallel execution of CPU and GPU threads.

\section{Odyssey}
\label{sec:odyssey}

The data series indexes discussed in the previous sections, 
operate on a single node. Thus, they do not  take advantage of the 
full computational power of modern distributed systems comprised of multiple multi-core nodes. 
Odyssey~\cite{odyssey} is a state-of-the-art {\em distributed} data-series processing framework,
which ensures good speedup and high scalability, thus addressing two major challenges 
of distributed data series indexing. 
An {\em ideal} distributed data series index should achieve linear speedup or be able to process
data collections whose size is proportional to the number of nodes in the system.
Experiments show~\cite{odyssey} that Odyssey does not notably depart from this ideal behavior. 

Odyssey provides a collection of scheduling schemes for query batches,
which result in good performance under several different workloads. 
These schemes are based on predicting the execution time of each query.
Odyssey provides a query analysis that shows correlation between the total execution 
time of a query and its initial best-so-far value. 
This analysis enables the calculation of the predictions. 

Odyssey achieves a good degree of load-balancing among the nodes
even in settings where the execution time predictions are not accurate.  
Odyssey's load balancing algorithm illustrates how to efficiently implement
the work-stealing approach~\cite{Cilk,AB98,10.1145/258492.258494,BL99,10.1145/378580.378639,FS99,FS00} in distributed data series indexing settings. 
In the work-stealing approach, 
nodes sitting idle may steal work from busy nodes.
The difficulty here is how to achieve this without ever moving any data
around, as moving data between nodes would be prohibitively expensive.
In Odyssey, this is esured by employing a (partial) replication scheme, which
is flexible enough to allow Odyssey to navigate through 
a fundamental trade-off between data scalability, and good performance during query answering. 
Specifically, no data replication would minimize space overhead,
but it would dissallow a cheap load-balancing solution, thus
resulting in higher execution times for answering queries. 
In Odyssey, a user may choose the replication degree 
that is the most relevant to its application based on its scalability and performance needs.

Odyssey is the first data series index that supports parallelization outside the boundaries of a single node,
without sacrificing the good performance of state-of-the-art indexes~\cite{peng2020messi,messijournal,peng2021sing}
within a node. 
This was not a trivial task, as experiments show that simple solutions of using as many instances of a state-of-the-art data series index 
as the number of nodes would not result in good performance, mainly due to severe load balancing problems. 
Supporting work-stealing on top of a state-of-the-art index would require moving data around.
Odyssey {\em single-node} indexing algorithm borrows techniques from MESSI~\cite{peng2020messi,messijournal}, but
it also provides new mechanisms to allow a node $v$ to rebuild parts of the index tree of another node $v'$
needed for executing the load that $v$ steals from $v'$. This requires the utilization of 
a different pattern of parallelism in traversing the index tree to produce the set of
data series that cannot be pruned, and new implementations for populating and processing the
data structures needed for efficient query answering. 

Experiments show that Odyssey's index creation exhibits perfect scalability as both the dataset size 
and the number of nodes increase.
Moreover, Odyssey exhibits up to 6.6x 
times faster exact query answering times than other state of the art 
parallel or distributed data series indexes.

\section{Discussion}
\label{sec:discussion}

We discussed state-of-the-art parallel and distributed indexes for answering similarity search queries
on big collections of data series in a fast and scalable way. We started with concurrency techniques
for disk-based indexes. We continued with efficient parallelization techniques for indexes 
built to work on in-memory data.
We also focused on techniques for data series processing that utilize GPUs to further speed up 
computation. Finally, we touched upon the main challenges that are introduced when moving from a single multi-core node 
to a distributed system with many multi-core nodes, and summarized state-of-the-art techniques for addressing these challenges. 

We focused on solving the problem of {\em exact} similarity search on big collections of {\em fixed-length} data series. 
Exact similarity search is a core operation needed in many critical data analytics tasks 
(including outlier detection, frequent pattern mining, 
clustering, classification, and others)~\cite{Palpanas2019}. 
An interesting open problem is whether the 
discussed techniques can be easily extended to efficiently support other types of similarity search queries,
such as approximate search, without or with (deterministic or probabilistic) guarantees~\cite{DBLP:journals/pvldb/EchihabiZPB19}. 
Also, it is interesting to study
how these parallelization techniques can be extended to indexes that handle data series 
of variable length~\cite{ulissejournal}.

Concurrent iSAX indexes are {\em locality-aware}. They maintain some form of data locality, 
and achieve high parallelism with low synchronization cost by having threads working, independently, 
on different parts of the data as much as possible, to avoid the need of frequent communication between threads.
However, they all employ locks, thus, they are blocking. If a thread holding a lock 
becomes slow (or crashes), the entire index blocks without being able to make any further progress. 
{\em Lock-freedom}~\cite{HS08} is a widely-studied property when designing
concurrent trees~\cite{ABF+22,DBLP:conf/podc/EllenFHR13,EFHR14,FR2018} 
and other data structures~\cite{FKR18,F04,HS08}. It avoids the use of locks, ensuring
that the system, as a whole, makes progress, independently
of delays (or failures) of threads. 
A recent study~\cite{fresh} presented the first lock-free iSAX index, FreSh, together
with Refresh, a generic approach that can be applied on top 
of any iSAX index to provide lock-freedom without adding any performance cost. 
An interesting open problem is to study the fundamental performance properties 
that govern other families of data series indexes, and come up with generic schemes for achieving 
locality-aware, lock-free synchronization on top of them.

The tree employed by an iSAX index has big root fan-out and its root subtrees are leaf-oriented binary trees. 
This type of trees support locality-aware parallelization, using locks, in a relatively easy way. 
Experimental work~\cite{EZPB18} has shown that no single indexing method is an overall winner 
in exact query answering. For this reason, recent indexing schemes~\cite{hercules} %for exact similarity search
incorporate key ideas from more than one data series indexing families %(and query answering algorithms)
with the goal of exhibiting the combined performance power of the different incorporated approaches. 
It would be interesting to see what kind of parallelization techniques would be
suitable to types of indexes that are not in the iSAX family~\cite{WP+13},
or for indexes that follow combined approaches~\cite{hercules}. 

Next generation computer systems will utilize emerging memory technologies, such as Non-Volatile Memory (NVM), 
to address the high computation demands of modern applications and provide persistence~\cite{ABF+22,FKK22}. 
The availability of non-volatile memory has increased the interest in the crash-recovery model, 
in which failed threads may be resurrected after the system crashes.
Developing a recoverable data series index that will exploit the NVM performance benefits 
and offer recoverability, distilling the best from both approaches, disk-based~\cite{parisplus} 
and in-memory~\cite{messi} data series indices, is an interesting open problem. 

\subsubsection{Acknowledgements} This work has been supported by 
the Hellenic Foundation for Research and Innovation (HFRI) under the ``Second Call for HFRI Research 
Projects to support Faculty Members and Researchers'' (project number: 3684).

\bibliographystyle{splncs04}
\bibliography{main}

\begin{thebibliography}{10}
\providecommand{\url}[1]{\texttt{#1}}
\providecommand{\urlprefix}{URL }
\providecommand{\doi}[1]{https://doi.org/#1}

\bibitem{Cilk}
Cilk: An efficient multithreaded runtime system. Journal of Parallel and
  Distributed Computing  \textbf{37}(1),  55--69 (1996)

\bibitem{Agrawal1993}
Agrawal, R., Faloutsos, C., Swami, A.N.: Efficient similarity search in
  sequence databases. In: FODO (1993)

\bibitem{AB98}
Arora, N.S., Blumofe, R.D., Plaxton, C.G.: Thread scheduling for
  multiprogrammed multiprocessors. In: Proceedings of the Tenth Annual ACM
  Symposium on Parallel Algorithms and Architectures. pp. 119--129. SPAA '98,
  Association for Computing Machinery, New York, NY, USA (1998).
  \doi{10.1145/277651.277678}, \url{https://doi.org/10.1145/277651.277678}

\bibitem{ABF+22}
Attiya, H., Ben-Baruch, O., Fatourou, P., Hendler, D., Kosmas, E.: Detectable
  recovery of lock-free data structures. In: Proceedings of the 27th ACM
  SIGPLAN Symposium on Principles and Practice of Parallel Programming. pp.
  262--277. PPoPP '22 (2022)

\bibitem{10.1145/258492.258494}
Blelloch, G.E., Gibbons, P.B., Matias, Y., Narlikar, G.J.: Space-efficient
  scheduling of parallelism with synchronization variables. In: Proceedings of
  the Ninth Annual ACM Symposium on Parallel Algorithms and Architectures. SPAA
  '97, Association for Computing Machinery (1997)

\bibitem{BL99}
Blumofe, R.D., Leiserson, C.E.: Scheduling multithreaded computations by work
  stealing. J. ACM  \textbf{46}(5),  720--748 (sep 1999).
  \doi{10.1145/324133.324234}, \url{https://doi.org/10.1145/324133.324234}

\bibitem{odyssey}
Chatzakis, M., Fatourou, P., Kosmas, E., Palpanas, T., Peng, B.: Odyssey: A
  journey in the land of distributed data series similarity search. Proc. VLDB
  Endow.  \textbf{16}(5),  1140--1153 (jan 2023).
  \doi{10.14778/3579075.3579087},
  \url{https://doi.org/10.14778/3579075.3579087}

\bibitem{hercules}
Echihabi, K., Fatourou, P., Zoumpatianos, K., Palpanas, T., Benbrahim, H.:
  {Hercules Against Data Series Similarity Search}. {PVLDB}  (2022)

\bibitem{bigdata20tutorial}
Echihabi, K., Zoumpatianos, K., Palpanas, T.: {Big Sequence Management: on
  Scalability (tutorial)}. In: {IEEE BigData} (2020)

\bibitem{EZPB18}
Echihabi, K., Zoumpatianos, K., Palpanas, T., Benbrahim, H.: The lernaean hydra
  of data series similarity search: An experimental evaluation of the state of
  the art. Proc. VLDB Endow.  \textbf{12}(2),  112--127 (oct 2018).
  \doi{10.14778/3282495.3282498},
  \url{https://doi.org/10.14778/3282495.3282498}

\bibitem{DBLP:journals/pvldb/EchihabiZPB19}
Echihabi, K., Zoumpatianos, K., Palpanas, T., Benbrahim, H.: Return of the
  lernaean hydra: Experimental evaluation of data series approximate similarity
  search. Proc. {VLDB} Endow.  \textbf{13}(3),  403--420 (2019).
  \doi{10.14778/3368289.3368303},
  \url{http://www.vldb.org/pvldb/vol13/p403-echihabi.pdf}

\bibitem{DBLP:conf/podc/EllenFHR13}
Ellen, F., Fatourou, P., Helga, J., Ruppert, E.: The amortized complexity of
  non-blocking binary search trees. In: {ACM} Symposium on Principles of
  Distributed Computing, {PODC} '14, Paris, France, July 15-18, 2014. pp.
  332--340 (2014). \doi{10.1145/2611462.2611486},
  \url{https://doi.org/10.1145/2611462.2611486}

\bibitem{EFHR14}
Ellen, F., Fatourou, P., Helga, J., Ruppert, E.: The amortized complexity of
  non-blocking binary search trees. In: Proc.\ 33rd ACM Symposium on Principles
  of Distributed Computing. pp. 332--340 (2014)

\bibitem{10.1145/378580.378639}
Fatourou, P.: Low-contention depth-first scheduling of parallel computations
  with write-once synchronization variables. In: Proceedings of the Thirteenth
  Annual ACM Symposium on Parallel Algorithms and Architectures. SPAA '01,
  Association for Computing Machinery, New York, NY, USA (2001).
  \doi{10.1145/378580.378639}, \url{https://doi.org/10.1145/378580.378639}

\bibitem{FKK22}
Fatourou, P., Kallimanis, N.D., Kosmas, E.: The performance power of software
  combining in persistence. In: Proceedings of the 27th ACM SIGPLAN Symposium
  on Principles and Practice of Parallel Programming. pp. 337--352. PPoPP '22,
  Association for Computing Machinery, New York, NY, USA (2022).
  \doi{10.1145/3503221.3508426}, \url{https://doi.org/10.1145/3503221.3508426}

\bibitem{FKR18}
Fatourou, P., Kallimanis, N.D., Ropars, T.: An efficient wait-free resizable
  hash table. In: Proceedings of the 30th on Symposium on Parallelism in
  Algorithms and Architectures. pp. 111--120. SPAA '18, Association for
  Computing Machinery, New York, NY, USA (2018). \doi{10.1145/3210377.3210408},
  \url{https://doi.org/10.1145/3210377.3210408}

\bibitem{fresh}
Fatourou, P., Kosmas, E., Palpanas, T., Paterakis, G.: Fresh: A lock-free data
  series index. In: International Symposium on Reliable Distributed Systems
  (SRDS) (2023)

\bibitem{FR2018}
Fatourou, P., Ruppert, E.: Persistent non-blocking binary search trees
  supporting wait-free range queries. CoRR  \textbf{abs/1805.04779} (2018),
  \url{http://arxiv.org/abs/1805.04779}

\bibitem{FS99}
Fatourou, P., Spirakis, P.: A new scheduling algorithm for general strict
  multithreaded computations. In: Jayanti, P. (ed.) Distributed Computing. pp.
  297--311. Springer Berlin Heidelberg, Berlin, Heidelberg (1999)

\bibitem{FS00}
Fatourou, P., Spirakis, P.: Efficient scheduling of strict multithreaded
  computations. Theory of Computing Systems  \textbf{33},  173--232 (2000)

\bibitem{Fekete:2016}
Fekete, J.D., Primet, R.: Progressive analytics: A computation paradigm for
  exploratory data analysis. CoRR  (2016)

\bibitem{F04}
Fraser, K.: Practical lock-freedom. Tech. Rep.~579, University of Cambridge
  Computer Laboratory (2004),
  \url{https://www.cl.cam.ac.uk/techreports/UCAM-CL-TR-579.pdf}

\bibitem{Airbus}
Guillaume, A.: {Head of Operational Intelligence Department Airbus. Personal
  communication.} (2017)

\bibitem{HS08}
Herlihy, M., Shavit, N.: The art of multiprocessor programming. Morgan Kaufmann
  (2008)

\bibitem{ulissejournal}
Linardi, M., Palpanas, T.: Scalable data series subsequence matching with
  ulisse. {VLDBJ}  (2020)

\bibitem{DBLP:journals/sigmod/Palpanas15}
Palpanas, T.: Data series management: The road to big sequence analytics.
  {SIGMOD} Record  (2015)

\bibitem{Palpanas2019}
Palpanas, T., Beckmann, V.: Report on the first and second interdisciplinary
  time series analysis workshop {(ITISA)}. SIGREC  (48(3), 2019)

\bibitem{messi}
Peng, B., Fatourou, P., Palpanas, T.: {MESSI: In-Memory Data Series Indexing}.
  In: {ICDE} (2020)

\bibitem{messijournal}
Peng, B., Fatourou, P., Palpanas, T.: Fast data series indexing for in-memory
  data. {VLDB} J.  \textbf{30}(6),  1041--1067 (2021)

\bibitem{peng2021sing}
Peng, B., Fatourou, P., Palpanas, T.: {SING: Sequence Indexing Using GPUs}. In:
  {Proceedings of the International Conference on Data Engineering (ICDE)}
  (2021)

\bibitem{sing}
Peng, B., Fatourou, P., Palpanas, T.: {SING:} sequence indexing using gpus. In:
  37th {IEEE} International Conference on Data Engineering, {ICDE} 2021,
  Chania, Greece, April 19-22, 2021. pp. 1883--1888. {IEEE} (2021).
  \doi{10.1109/ICDE51399.2021.00171},
  \url{https://doi.org/10.1109/ICDE51399.2021.00171}

\bibitem{peng2018paris}
Peng, B., Palpanas, T., Fatourou, P.: Paris: The next destination for series
  indexing and query answering. IEEE BigData  (2018)

\bibitem{peng2020messi}
Peng, B., Palpanas, T., Fatourou, P.: Messi: In-memory data series indexing.
  In: ICDE (2020)

\bibitem{parisplus}
Peng, B., Palpanas, T., Fatourou, P.: Paris+: Data series indexing on
  multi-core architectures. TKDE  (2020)

\bibitem{rakthanmanon2012searching}
Rakthanmanon, T., Campana, B.J.L., Mueen, A., Batista, G.E.A.P.A., Westover,
  M.B., Zhu, Q., Zakaria, J., Keogh, E.J.: Searching and mining trillions of
  time series subsequences under dynamic time warping. In: SIGKDD (2012)

\bibitem{iSAX}
Shieh, J., Keogh, E.: Isax: Indexing and mining terabyte sized time series
  (2008). \doi{10.1145/1401890.1401966},
  \url{https://doi.org/10.1145/1401890.1401966}

\bibitem{WP+13}
Wang, Y., Wang, P., Pei, J., Wang, W., Huang, S.: A data-adaptive and dynamic
  segmentation index for whole matching on time series. Proc. VLDB Endow.
  \textbf{6}(10),  793--804 (aug 2013). \doi{10.14778/2536206.2536208},
  \url{https://doi.org/10.14778/2536206.2536208}

\bibitem{zoumpatianos2016ads}
Zoumpatianos, K., Idreos, S., Palpanas, T.: Ads: the adaptive data series
  index. {VLDB} J.  (2016)

\bibitem{ZoumpatianosIP16}
Zoumpatianos, K., Idreos, S., Palpanas, T.: {ADS:} the adaptive data series
  index. {VLDB} J.  (2016)

\bibitem{fulfillingtheneed}
Zoumpatianos, K., Palpanas, T.: Data series management: Fulfilling the need for
  big sequence analytics. In: ICDE (2018)

\end{thebibliography}
\end{document}